%% file: main.tex
\documentclass[sigconf]{acmart}
\RequirePackage{etex}
\usepackage{booktabs} 
\usepackage[normalem]{ulem}
\usepackage{amsmath,array}
\usepackage{listings}
\usepackage{graphicx}
\usepackage{caption}
\usepackage{subfig}
\usepackage{standalone}
 \usepackage{multirow}
\usepackage{tikz}
\usetikzlibrary{shapes,arrows,positioning,fit,matrix}
\usepackage{color}
\usepackage[colorinlistoftodos]{todonotes}
\usepackage{mathtools}
\usepackage{mdframed}
\usepackage{listings}
\usepackage{color}
\usepackage{url}
\usepackage{balance}

\graphicspath{{Figures/}}
\DeclareGraphicsExtensions{.pdf,.jpeg,.png}

\definecolor{Gray}{gray}{0.9}
\definecolor{codeBG}{rgb}{0.88,1,1}
\definecolor{codeBG2}{rgb}{1,0.9,0.7}

\newcommand{\transfName}{inspector-guided }
\newcommand{\TransfName}{Inspector-guided }

\setcopyright{rightsretained}

\acmDOI{10.475/123_4}

\acmISBN{123-4567-24-567/08/06}

\acmConference[]{}{}{}
\acmYear{2017}

\begin{document}

\title[Sympiler: Transforming Sparse Codes]{Sympiler: Transforming Sparse Matrix Codes by Decoupling Symbolic Analysis}

\author{Kazem Cheshmi}
\orcid{1234-5678-9012}
\affiliation{%
  \institution{Rutgers University}
  \city{Piscataway} 
  \state{NJ} 
  \country{US}
}
\email{kazem.ch@rutgers.edu}

\author{Shoaib Kamil}
\affiliation{%
  \institution{Adobe Research}
  \city{Cambridge} 
  \state{MA} 
  \country{US}
}
\email{kamil@adobe.com}

\author{Michelle Mills Strout}
\affiliation{%
  \institution{University of Arizona}
  \city{Tucson} 
  \state{AZ}
  \country{US}
  }
\email{mstrout@cs.arizona.edu}

\author{Maryam Mehri Dehnavi}
\affiliation{
  \institution{Rutgers University}
    \city{Piscataway}
  \state{NJ}
  \country{US}
  }
\email{maryam.mehri@rutgers.edu}


\begin{abstract}
Sympiler is a domain-specific code generator that optimizes sparse matrix computations by decoupling the symbolic analysis phase from the numerical manipulation stage in sparse codes. The computation patterns in sparse numerical methods are guided by the input sparsity structure and the sparse algorithm itself. In many real-world simulations, the sparsity pattern changes little or not at all. Sympiler takes advantage of these properties to symbolically analyze sparse codes at compile-time and to apply inspector-guided transformations that enable applying low-level transformations to sparse codes. As a result, the Sympiler-generated code outperforms highly-optimized matrix factorization codes from commonly-used specialized libraries, obtaining average speedups over Eigen and CHOLMOD of 3.8$\times$ and 1.5$\times$ respectively.
\end{abstract}

%
%
%



\settopmatter{printacmref=false} 
\renewcommand\footnotetextcopyrightpermission[1]{} 
\maketitle
\section{Introduction}
\input{SRC/Introduction}
\section{Sympiler: A Symbolic-Enabled Code Generator }
\input{SRC/CompilerInternal}

\section{Case studies}
\input{SRC/CaseStudy}


\section{Experimental results}
\input{SRC/Results}

\section{Related Work}
\input{SRC/RelatedWork}

\section{Conclusion}

In this paper we demonstrated how decoupling symbolic analysis from numerical manipulation can enable the generation of domain-specific highly-optimized sparse codes with static sparsity patterns. Sympiler, the proposed domain-specific code generator, takes the sparse matrix pattern and the sparse matrix algorithm as inputs to perform symbolic analysis at compile-time. It then uses the information from symbolic analysis to apply a number of inspector-guided and low-level transformations to the sparse code. The Sympiler-generated code outperforms two state-of-the-art sparse libraries, Eigen and CHOLMOD, for the sparse Cholesky and the sparse triangular solve algorithms.

\bibliographystyle{ACM-Reference-Format}
\bibliography{sigproc}

\end{document}

%% file: SRC/Introduction.tex
\begin{figure*}[!ht]
	\centering
	\begin{tabular}[b]{ccc}

	\multicolumn{2}{c}{\subfloat[\label{fig:reachset}]{
    \includegraphics[width=0.65\textwidth]{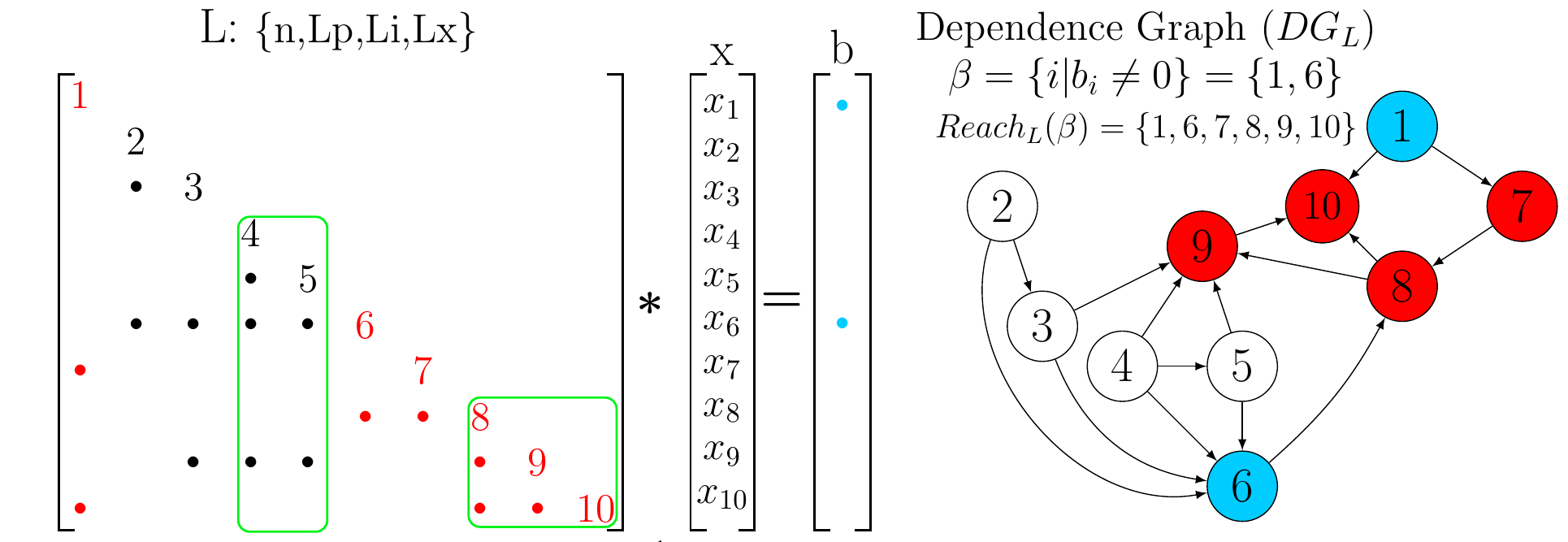}}}
	&
	\multirow{2}{*}[1.95in]{
	\begin{tabular}{r}
	\renewcommand{\thesubfigure}{b}
	\subfloat[\label{fig:forward} Forward substitution]{\includegraphics[width=0.3\textwidth]{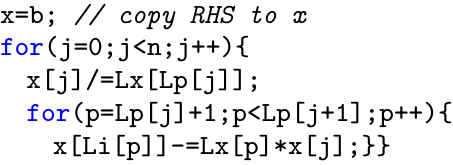}}
	\\
	\renewcommand{\thesubfigure}{e}
	\subfloat[\label{fig:unrolled} Sympiler-generated]{\includegraphics[width=0.3\textwidth]{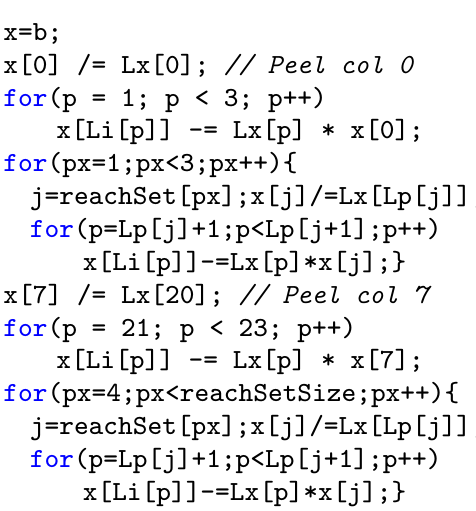}}
	\end{tabular}
	}
	\\
    \\
	\renewcommand{\thesubfigure}{c} \subfloat[\label{fig:library} Library implementation]{\includegraphics[width=0.3\textwidth]{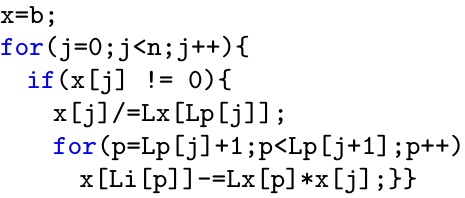}}
	&\renewcommand{\thesubfigure}{d} \subfloat[\label{fig:dec}Decoupled code]{\includegraphics[width=0.3\textwidth]{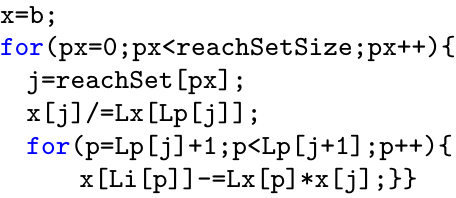}}
	\end{tabular}
\caption{\label{fig:Motive}%
  Four different codes for solving the linear system in \protect\subref{fig:reachset}. In all four code variants, the matrix $L$ is stored in compressed sparse column (CSC) format, with \{\texttt{n,Lp,Li,Lx}\} representing \{matrix order, column pointer, row index, nonzeros\} respectively. The dependence graph $DG_L$ is the adjacency graph of matrix $L$; the vertices of $DG_L$ correspond to columns of $L$ and its edges show dependencies between columns in triangular solve. Vertices corresponding to nonzero columns are colored blue, and columns that must participate in the computation due to dependence structure are colored red; the white vertices can be skipped during computation. The boxes around columns show supernodes of different sizes. \protect\subref{fig:forward} is a forward substitution algorithm. \protect\subref{fig:library} is a library implementation that skips iterations when the corresponding entry in the $x$ is zero. \protect\subref{fig:dec} is the decoupled code that uses symbolic information given in $\mathit{reachSet}$, which is computed by performing a depth-first search on $DG_L$. \protect\subref{fig:unrolled} is the Sympiler-generated code which peels iterations corresponding to columns within the reach-set with more than 2 nonzeros.
  }
\end{figure*}

Sparse matrix computations are at the heart of many scientific applications and data analytics codes.  The performance and efficient memory usage of these codes depends heavily on their use of specialized sparse matrix data structures that only store the nonzero entries.  However, such compaction is done using index arrays that result in indirect array accesses. Due to these indirect array accesses, it is difficult to apply conventional compiler optimizations such as tiling and vectorization even for static index array operations like sparse matrix vector multiply. A static index array does not change during the algorithm; for more complex operations with dynamic index arrays such as matrix factorization and decomposition, the nonzero structure is modified during the computation, making conventional compiler optimization approaches even more difficult to apply.

The most common approach to accelerating sparse matrix computations is to identify a specialized library that provides a manually-tuned implementation of the specific sparse matrix routine.  A large number of sparse libraries are available (e.g., SuperLU \cite{SUdemmel1999supernodal}, MUMPS \cite{MUMPS1amestoy2000multifrontal}, CHOLMOD \cite{CHOLMODchen2008algorithm}, KLU \cite{KLUdavis2010algorithm}, UMFPACK \cite{SuiteSparseQRdavis2011algorithm}) for different numerical kernels, supported architectures, and specific kinds of matrices.  While hand-written specialized libraries can provide high performance, they must be manually ported to new architectures and may stagnate as architectural advances continue.  Alternatively, compilers can be used to optimize code while providing architecture portability.  
However, indirect accesses and the resulting complex dependence structure run into compile-time loop transformation framework limitations.

Compiler loop transformation frameworks such as those based on the polyhedral model use algebraic representations of loop nests to transform code and successfully generate highly-efficient dense matrix kernels ~\cite{ancourt1991scanning, kelly1998optimization, quillere2000generation, vasilache2006polyhedral, tiwari2009scalable,chen2012polyhedra}.  However, such frameworks are limited when dealing with non-affine loop bounds and/or array subscripts, both of which arise in sparse codes.  
Recent work has extended polyhedral methods to effectively operate on kernels with static index arrays by building run-time \textit{inspectors} that examine the nonzero structure and \textit{executors} that use this knowledge to transform code execution~\cite{venkat2015loop, strout2016approach, van2010sublimation, venkat2014non, venkat2016automating}.
However, these techniques are limited to transforming sparse kernels with static index arrays.  
Sympiler addresses these limitations by performing \textit{symbolic analysis} at compile-time to compute fill-in structure and to remove dynamic index arrays from sparse matrix computations. Symbolic analysis is a term from the numerical computing community.  It refers to phases that determine the computational patterns that only depend
on the nonzero pattern and not on numerical values. Information from symbolic analysis can be used to make subsequent numeric manipulation faster, and the information can be reused as long as the matrix nonzero structure remains constant.

For a number of sparse matrix methods such as LU and Cholesky, it is well known that viewing their computations as a graph (e.g., elimination tree, dependence graph, and quotient graph) and applying a method-dependent graph algorithm yields information about dependences that can then be used to more efficiently compute the numerical method ~\cite{davis2006direct}. 
Most high-performance sparse matrix computation libraries utilize symbolic information, but couple this symbolic analysis with numeric computation, further making it difficult for compilers to optimize such codes.

This work presents Sympiler, which generates high-performance sparse matrix code by fully decoupling the symbolic analysis from numeric computation and transforming code to utilize the symbolic information.  After obtaining symbolic information by running a symbolic inspector, Sympiler applies \transfName transformations, such as variable-sized blocking, resulting in performance equivalent to hand-tuned libraries. But Sympiler goes further than existing numerical libraries by generating code for a specific matrix nonzero structure.
Because matrix structure often arises from properties of the underlying physical system that the matrix represents, in many cases the same structure reoccurs multiple times, with different values of nonzeros.  Thus, Sympiler-generated code can combine \transfName and low-level transformations to produce even more efficient code. The transformations applied by Sympiler improves the performance of sparse matrix codes through applying optimizations for a single-core such as vectorization and increased data locality which should extend to improve performance on shared and distributed memory systems.

\subsection{Motivating Scenario}
\label{sec:introMot}
Sparse triangular solve takes a lower triangular matrix $L$ and a right-hand side (RHS) vector $b$ and solves the linear equation $Lx=b$ for $x$. It is a fundamental building block in many numerical algorithms such as factorization \cite{davis2006direct,li2005overview}, direct system solvers \cite{LDLdavis2005algorithm}, and rank update methods \cite{davis2009dynamic}, where the RHS vector is often sparse. A na\"ive implementation visits every column of matrix $L$ to propagate the contributions of its corresponding $x$ value to the rest of $x$ (see Figure~\ref{fig:forward}).  However, with a sparse $b$, the solution vector is also sparse, reducing the required iteration space of sparse triangular solve to be proportional to the number of nonzero values in $x$. Taking advantage of this property requires first determining the nonzero pattern of $x$. Based on a theorem from Gilbert and Peierls \cite{gilbert1988sparse}, the  \textit{dependence graph} $DG_L = (V, E)$ for matrix $L$ with nodes $V=\{1,...,n\}$ and edges $E=\{(j, i) | L_{ij} \neq 0\}$ can be used to compute the  nonzero pattern of $x$, where $n$ is the matrix rank and numerical cancellation is neglected. The nonzero indices in $x$ are given by $Reach_L(\beta)$ which is the set of all nodes reachable from any node in $\beta=\{i|b_i\neq{0}\}$, and can be computed with a depth-first search of the directed graph $DG_L$ staring with $\beta$. 
An example dependence graph is illustrated in Figure~\ref{fig:reachset}. The blue colored nodes correspond to set $\beta$ and the final \textit{reach-set} $Reach_L(\beta)$ contains all the colored nodes in the graph. 

Figure~\ref{fig:Motive} shows four different implementations of sparse triangular solve. 
Most solvers assume the input matrix \textit{L} is stored in a compressed sparse column (CSC) storage format.  While the na\"ive implementation in Figure~\ref{fig:forward} traverses all columns, the typical library implementation shown in Figure~\ref{fig:library} skips iterations when the corresponding value in \textit{x} is zero. 

The implementation in Figure~\ref{fig:dec} shows a decoupled code that uses the symbolic information provided by the pre-computed reach-set. This decoupling simplifies numerical manipulation and reduces the run-time complexity from $O(|b|+n+f)$ in Figure~\ref{fig:library} to $O(|b|+f)$ in Figure~\ref{fig:dec}, where $f$ is the number of floating point operations and $|b|$ is the number of nonzeros in $b$. Sympiler goes further by building the reach-set at compile-time and leveraging it to generate code specialized for the specific matrix structure and RHS. The Sympiler-generated code is shown in Figure~\ref{fig:unrolled}, where the code only iterates over reached columns and peels iterations where the number of nonzeros in a column is greater than some threshold (in the case of the figure, this threshold is 2). These peeled loops can be further transformed with vectorization to speed up execution.  This shows the power of fully decoupling the symbolic analysis phase from the code that manipulates numeric values: the compiler can aggressively apply conventional optimizations, using the reach-set to guide the transformation. On matrices from the SuiteSparse Matrix Collection, the Sympiler-generated code shows speedups between 8.4$\times$ to 19$\times$ with an average of 13.6$\times$ compared to the forward solve code (Figure~\ref{fig:forward}) and from 1.2$\times$ to 1.7$\times$ with an average of 1.3$\times$ compared to the library-equivalent code (Figure~\ref{fig:library}).

\subsection{Static Sparsity Patterns}
A fundamental concept that Sympiler is built on is that the structure of sparse matrices in scientific codes is dictated by the physical domain and as such does not change in many applications. For example, in power system modeling and circuit simulation problems the sparse matrix used in the matrix computations is often a Jacobian matrix, where the structure is derived from interconnections among the power system and circuit components such as generation, transmission, and distribution resources. While the numerical values in the sparse input matrix change often, a change in the sparsity structure occurs on rare occasions with a change in circuit breakers, transmission lines, or one of the physical components. The sparse systems in simulations in domains such as electromagentics \cite{luna2013packaged, dorf2006electronics, exposito2012dual,ElMehri2015}, computer graphics \cite{gibson1997survey}, and fluid mechanics \cite{anderson1984computational} are assembled by discretizing a physical domain and approximating a partial differential equation on the mesh elements. A sparse matrix method is then used to solve the assembled systems. The sparse structure originates from the physical discretization and therefore the sparsity pattern remains the same except where there are deformations or if adaptive mesh refinement is used. Sparse matrices in many other physical domains exhibit the same behavior and benefit from Sympiler.

\subsection{Contributions}
This work describes Sympiler, a sparsity-aware code generator for sparse matrix algorithms that leverages symbolic information to generate fast code for a specific matrix structure.  The major contributions of this paper are:
\begin{itemize}
  \item A novel approach for building \textit{compile-time symbolic inspectors} that obtain information about a sparse matrix, to be used during compilation.
  \item \textit{\TransfName transformations} that leverage compile-time information to transform sparse matrix code for specific algorithms.
  \item Implementations of symbolic inspectors and \transfName transformations for two algorithms: sparse triangular solve and sparse Cholesky factorization.
  \item A demonstration of the performance impact of our code generator, showing that Sympiler-generated code can outperform state-of-the-art libraries for triangular solve and Cholesky factorization by up to 1.7$\times$ and 6.3$\times$ respectively.
\end{itemize}

%% file: SRC/CompilerInternal.tex
\label{sec:transformation}
\begin{figure*}[ht]
  \centering
  \includegraphics[width=\textwidth]{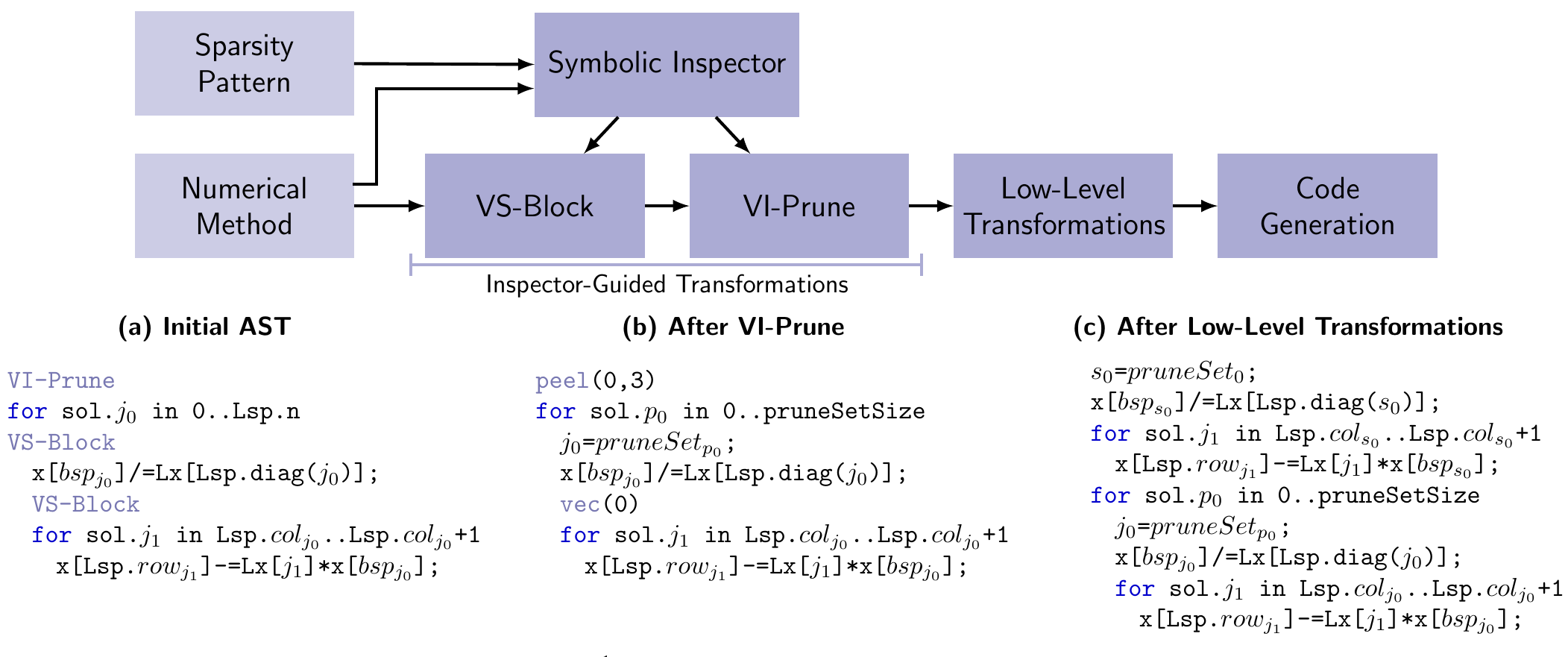}
  \caption{Sympiler lowers a functional representation of a sparse kernel to imperative code using the inspection sets. 
  Sympiler constructs a set of loop nests and annotates them with some domain-specific information that is later used in \transfName transformations. The \transfName transformations use the lowered code and inspection sets as input and apply transformations. \TransfName transformations also provide hints for further low-level transformations by annotating the code. For instance, the transformation steps for the code in \protect\autoref{fig:Motive} are: (a) Initial AST with annotated information showing where the VI-Prune and VS-Block transformations apply. (b) The symbolic inspector sends the reach-set as \texttt{pruneSet}, which VI-Prune uses to add hints to further steps--- in this case, peeling iterations 0 and 3. (c). The hinted low-level transformations are applied and final code generated (peeling is only shown for the iteration zero).}
  \label{fig:Compiler-Overview}
\end{figure*}

Sympiler generates efficient sparse kernels by tailoring sparse code to specific matrix sparsity structures. By decoupling the symbolic analysis phase, Sympiler uses information from symbolic analysis to guide code generation for the numerical manipulation phase of the kernel.  
In this section, we describe the overall structure of the Sympiler code generator, as well as the domain-specific transformations enabled by leveraging information from the symbolic inspector.

\subsection{Sympiler Overview}
\label{sec:syminputs}
Sparse triangular solve and Cholesky factorization are currently
implemented in Sympiler.  Given one of these numerical methods and an input matrix stored using compressed sparse column (CSC) format, Sympiler utilizes a method-specific \textit{symbolic inspector} to obtain information about the matrix.  This information is used to apply domain-specific optimizations while lowering the code for the numerical method. In addition, the lowered code is annotated with additional low-level transformations (such as unrolling) when applicable based on domain- and matrix-specific information.  Finally, the annotated code is further lowered to apply low-level optimizations and output to C source code.

Code implementing the numerical solver is represented in a domain-specific abstract syntax tree (AST). Sympiler produces the final code by applying a series of phases to this AST, transforming the code in each phase.  An overview of the process is shown in Figure~\ref{fig:Compiler-Overview}.  The initial AST for triangular solve is shown in Figure~\ref{fig:Compiler-Overview}a prior to any transformations.

\begin{figure*}[!t]
\begin{tabular}{cc}
\centering
&\\
\hline
  \subfloat{\input{Figures/pruneBefore.tex}
\label{fig:pruneBefore}} 
  &\subfloat{\input{Figures/pruneAfter.tex}
\label{fig:prunekAfter}}\\
  {\footnotesize\textbf{(a) Before}} & {\footnotesize\textbf{(b) After}}\\
\multicolumn{2}{c}{\footnotesize \textbf{Variable Iteration Space Pruning}, loop[k].VI-Prune(pruneSet,pruneSetSize)} \\
\hline
  \subfloat{\input{Figures/block.tex}
\label{fig:blockBefore}}
  &\subfloat{\input{Figures/blockAfter.tex}
\label{fig:blockAfter}} \\
  {\footnotesize\textbf{(c) Before}} & {\footnotesize\textbf{(d) After}}\\
\multicolumn{2}{c}{\footnotesize \textbf{2D Variable-Sized Blocking}, loop[I].VS-Block(blockSet,blockSetSize)} \\
\end{tabular}
  \caption{The \transfName transformations. \textbf{Top:} The loop over $I_k$ with iteration space $m$ in \protect\subref{fig:pruneBefore} transforms to a loop  over $I_p$ with iteration space $\mathit{pruneSetSize}$ in \protect\subref{fig:prunekAfter}.  Any use of the original loop index $I_k$ is replaced with its corresponding value from $\mathit{pruneSet}$ i.e., $I'_k$. \textbf{Bottom:} The two nested loops in \protect\subref{fig:blockBefore} are transformed into loops over variable-sized blocks in \protect\subref{fig:blockAfter}. 
  }
  
\label{fig:TransformOverall}
\end{figure*}
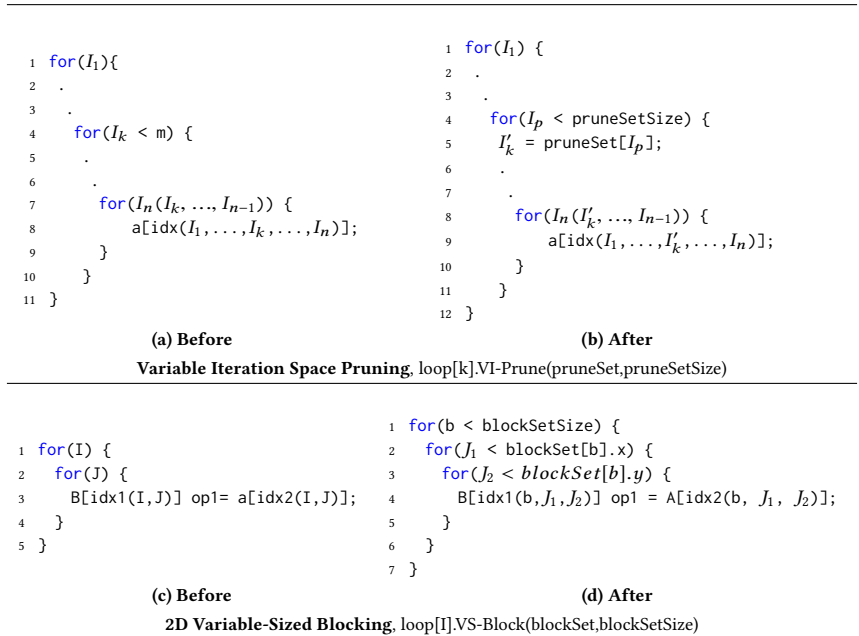

\subsection{Symbolic Inspector}
\label{sec:symInsp}
Different numerical algorithms can make use of symbolic information in different ways, and prior work has described run-time graph traversal strategies for various numerical methods~\cite{pothen2004elimination,SUDISTli2003superlu_dist,UMFPACKdavis2004algorithm,davis2006direct}. The compile-time inspectors in Sympiler are based on these strategies.
For each class of numerical algorithms with the same symbolic analysis approach, Sympiler uses a specific symbolic inspector to obtain information about the sparsity structure of the input matrix and stores it in an algorithm-specific way for use during later transformation stages.  

We classify the used symbolic inspectors based on the numerical method as well as the transformations enabled by the obtained information.  For each combination of algorithm and transformation, the symbolic inspector creates an \textit{inspection graph} from the given sparsity pattern and traverses it during inspection using a specific \textit{inspection strategy}. The result of the inspection is the \textit{inspection set}, which contains the result of running the inspector on the inspection graph.  Inspection sets are used to guide the transformations in Sympiler. Additional numerical algorithms and transformations can be added to Sympiler, as long as the required inspectors can be described in this manner as well. 

For our motivating example, triangular solve, the \textit{reach-set} can be used to prune loop iterations that perform work that is unnecessary due to the sparseness of matrix or the right hand side.  In this case, the inspection set is the reach-set, and the inspection strategy is to perform a depth-first search over the inspection graph, which is the directed dependency graph $DG_L$ of the triangular matrix.  For the example linear system shown in Figure~\ref{fig:Motive}, the symbolic inspector generates the reach-set \{6, 1, 7, 8, 9, 10\}.

\subsection{\TransfName Transformations}
\label{sec:algoTrans}
The initial lowered code along with the inspection sets obtained by the symbolic inspector are passed to a series of passes that further transform the code. Sympiler currently supports two transformations guided by the inspection sets: \textit{Variable Iteration Space Pruning} and \textit{2D Variable-Sized Blocking}, which can be applied independently or jointly depending on the input sparsity. As shown in Figure~\ref{fig:Compiler-Overview}a, the code is annotated with information showing where \transfName transformations may be applied. The symbolic inspector provides the required information to the transformation phases, which decide whether to transform the code based on the inspection sets. Given the inspection set and annotated code, transformations occur as illustrated in Figure~\ref{fig:TransformOverall}.

\subsubsection{Variable Iteration Space Pruning}
\label{subsec:VItrans}

 Variable Iteration Space Pruning (VI-Prune) prunes the iteration space of a loop using information about the sparse computation. The iteration space for sparse codes can be considerably smaller than that for dense codes, since the computation needs to only consider iterations with nonzeros. The inspection stage of Sympiler generates an inspection set that enables transforming the unoptimized sparse code to a code with a reduced iteration space.

Given this inspection set, the VI-Prune transformation can be applied at a particular loop-level to the sparse code to transform it from Figure~\ref{fig:pruneBefore} to Figure~\ref{fig:prunekAfter}.  In the figure, the transformation is applied to the $k^{th}$ loop nest in line 4. In the transformed code the iteration space is pruned to \texttt{pruneSetSize}, which is the inspection set size. In addition to the new loop, all references to $I_k$ (the loop index before transformation) are replaced by its corresponding value from the inspection set, \texttt{pruneSet[$I_p$]}.  Furthermore, the transformation phase utilizes inspection set information to annotate specific loops with further low-level optimizations to be applied by later stages of code generation.  These annotations are guided by thresholds that decide when specific low-level optimizations result in faster code.

In our running example of triangular solve, the generated inspection set from the symbolic inspector enables reducing the iteration space of the code. The VI-Prune transformation elides unnecessary iterations due to zeros in the right hand side.  In addition, depending on the number of iterations the loops will run (which is known thanks to the symbolic inspector), loops are annotated with directives to unroll and/or vectorize during code generation.

\subsubsection{2D Variable-Sized Blocking}
\label{sec:block}
2D Variable-Sized Blocking (VS-Block) converts a sparse code to a set of non-uniform dense sub-kernels. In contrast to the conventional approach of blocking/tiling dense codes, where the input and computations are blocked into smaller uniform sub-kernels, the unstructured computations and inputs in sparse kernels make blocking optimizations challenging. The symbolic inspector identifies sub-kernels with similar structure in the sparse matrix methods and the sparse inputs to provide the VS-Block stage with ``blockable'' sets that are not
necessarily of the same size or consecutively located. These blocks are similar to the concept of \textit{supernodes}~\cite{li2005overview} in sparse libraries. VS-Block must deal with a number of challenges:
\begin{itemize}
\item The block sizes are variable in a sparse kernel.
\item Due to using compressed storage formats, the block elements may not be in consecutive memory locations.
\item The type of numerical method used may need to change after applying this transformation. For example, applying VS-Block to Cholesky factorization requires dense Cholesky factorization on the diagonal segment of the blocks, and the off-diagonal segments of the blocks must be updated using a set of dense triangular solves.
\end{itemize}

 To address the first challenge, the symbolic inspector uses an inspection strategy that provides an inspection set specifying the size of each block. For the second challenge, the transformed code allocates temporary block storage and copies data as needed prior to operating on the block. Finally, to deal with the last challenge, the synthesized loops/instructions in the lowering phase contain information about the block location in the matrix, and when applying this transformation, the correct operation is chosen for each loop/instruction. As with the VI-Prune transformation, VS-Block also annotates loops with further low-level transformations such as tiling to be applied during code generation. By leveraging specific information about the matrix when applying the transformation, Sympiler is able to mitigate all of the difficulties of applying VS-Block to sparse numerical methods.  

An off-diagonal version of the VS-Block transformation is shown in Figures~\ref{fig:blockBefore} and~\ref{fig:blockAfter}. As shown, a new outer loop is made that provides the block information to the inner loops using the given $\mathit{blockSet}$. 
The inner loop in Figure~\ref{fig:blockBefore} transforms to two nested loops (lines 2--6) that iterate over the block specified by the outer loop. The diagonal version VS-Block heavily depends on domain information. More detailed examples of applying this transformation to triangular solve and Cholesky factorization is described in Section~\ref{sec:caseStudy}.


\subsection{Enabled Conventional Low-level Transformations}
\label{sec:enabledTrans}
While applying \transfName transformations, the original loop nests are transformed into new loops with potentially different iteration spaces, enabling the application of conventional low-level transformations. Based on the applied \transfName transformations as well as the properties of the input matrix and right-hand side vectors, the code is annotated with some transformation directives. An example of these annotations are shown in Figure~\ref{fig:Compiler-Overview}b where loop peeling is annotated within the VI-Pruned code. To decide when to add these annotations, the \transfName transformations use sparsity-related parameters such as the average block size. The main sources of enabling low-level transformations are:
\begin{enumerate}
\item Symbolic information provides dependency information at compile-time, allowing Sympiler to apply more transformations such as peeling based on the reach-set in Figure~\ref{fig:Motive}; 
\item \TransfName transformations remove some of the indirect memory accesses and annotate the code with potential conventional transformations;
\item Sparsity-specific code generation enables Sympiler to know details such as loop boundaries at compile-time. Thus, several customized transformations 
 are applied such vectorization of loops with iteration counts greater than a threshold; 
\end{enumerate}

Figure~\ref{fig:unrolled} shows how some of the iterations in the triangular solve code after VI-Prune can be peeled. In this example, the inspection set used for VI-Prune is the reach-set $\{1,6,8,9,10\}$. Because the reach-set is created in topological order, iteration ordering dependencies are met and thus code correctness is guaranteed after loop peeling. As shown in \autoref{fig:Compiler-Overview}b, the transformed code after VI-Prune is annotated with the enabled peeling transformation based on the number of nonzeros in the columns (the \textit{column count}). The two selected iterations with column count greater than 2 are peeled to replace them with a specialized kernel or to apply another transformation such as vectorization. 

%% file: Figures/pruneBefore.tex
\definecolor{codeBG3}{rgb}{0.88,1,1}
\newcommand{\cL}{{\cal L}}
\input{Figures/codestyle.tex}
\begin{lstlisting}[mathescape=true,frame=none]
for($I_1$){
 .
  .
   for($I_k$ < m) { |\label{line:pLoop}  |     
    .
     .
      for($I_n$($I_k,..., I_{n-1}$)) {
          a[idx($I_1$,...,$I_k$,...,$I_n$)];                                                      
      }
    }
}
  \end{lstlisting}


%% file: Figures/pruneAfter.tex
\input{Figures/codestyle.tex}

\begin{lstlisting}[mathescape=true,frame=none]
for($I_1$) {
 .
  .
   for($I_p$ < pruneSetSize) { |\label{line:pLoopAfter}|
    $I'_k$ = pruneSet[$I_p$];
    .
     .
      for($I_n$($I'_k,..., I_{n-1}$)) {
          a[idx($I_1$,...,$I'_k$,...,$I_n$)];                                                      
      }
    }
}
\end{lstlisting}


%% file: Figures/block.tex
\newcommand{\cL}{{\cal L}}
\input{Figures/codestyle.tex}
\begin{lstlisting}[mathescape=true,frame=none]
for(I) {
  for(J) {
    B[idx1(I,J)] op1= a[idx2(I,J)];
  }        
}
  \end{lstlisting}


%% file: Figures/blockAfter.tex
\newcommand{\cL}{{\cal L}}
\input{Figures/codestyle.tex}
\begin{lstlisting}[mathescape=true,frame=none]
for(b < blockSetSize) {
  for($J_1$ < blockSet[b].x) {
    for($J_2 < blockSet[b].y$) {
      B[idx1(b,$J_1$,$J_2$)] op1 = A[idx2(b, $J_1$, $J_2$)];
    }
  }
}
  \end{lstlisting}


%% file: SRC/CaseStudy.tex
\label{sec:caseStudy}

\begin{figure}[ht]
  \centering
  
  \includegraphics[width=.48\textwidth]{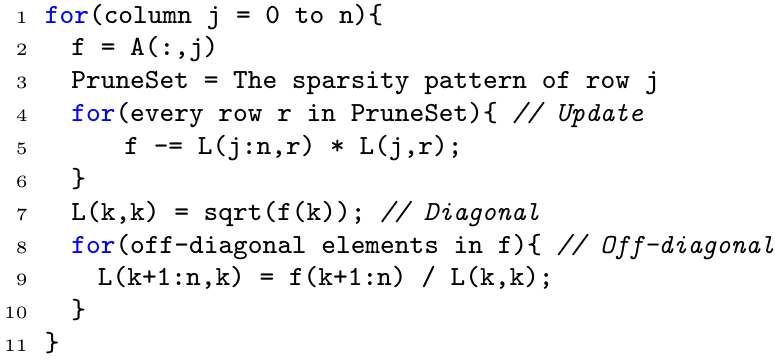}
  \caption{The pseudo-code of the left-looking Cholesky. }
  \label{fig:codeCholesky}
  
\end{figure}

\begin{table*}[ht]

\centering
\caption{Inspection and transformation elements in Sympiler for triangular solve and Cholesky. DG: dependency graph, SP (RHS): sparsity patterns of the right-hand side vector, DFS: depth-first search, SP(A): sparsity patterns of the coefficient $A$, SP ($L_j$): sparsity patterns of the $j^{th}$row of $L$, unroll: loop unrolling, peel: loop peeling, dist: loop distribution, tile: loop tiling.}
\begin{tabular}{|p{2.4cm}|p{1.6cm}|p{1.6cm}|p{1.8cm}|p{1.6cm}|p{1.6cm}|p{1.8cm}|p{2.2cm}|} \hline
\multirow{2}{*}{\textbf{Transformations}} &\multicolumn{3}{c|}{\textbf{Triangular Solve}}&\multicolumn{3}{c|}{ \textbf{Cholesky}}&\\ 
  &Inspection Graph & Inspection Strategy & Inspection Set & Inspection Graph  & Inspection Strategy & Inspection Set& Enabled \mbox{Low-level}\\ \hline
\textbf{VI-Prune} & DG + & DFS& Prune-set & etree + & Single-node & Prune-set & dist, unroll, peel, \\ 
  & SP(RHS) & & (reach-set) & SP($A$) & up-traversal & (SP($L_j$)) & vectorization\\\hline
  \textbf{VS-Block} & DG& Node \mbox{equivalence} & Block-set \mbox{(supernodes)} & etree~+ \mbox{ColCount($A$)} & \mbox{Up-traversal} &Block-set \mbox{(supernodes)} &  tile, unroll, peel, vectorization  \\ 
\hline\end{tabular}
\label{tab:inspection}
\end{table*}

Sympiler currently supports two important sparse matrix computations: triangular solve and Cholesky factorization.
 This section discusses some of the graph theory and algorithms used in Sympiler's symbolic inspector to extract inspections sets for these two matrix methods. The run-time complexity of the Symbolic inspector is also presented to evaluate inspection overheads. Finally, we demonstrate how the VI-Prune and VS-Block transformations are applied using the inspection sets. 
 Table~\ref{tab:inspection} shows a classification of the inspection graphs, strategies, and resulting inspection sets for the two studied numerical algorithms in Sympiler. As shown in Table~\ref{tab:inspection}, the symbolic inspector performs a set of known inspection methods and generates some sets which includes symbolic information. The last column of Table~\ref{tab:inspection} shows the list of transformations enabled by each \transfName transformation. 
 We also discuss extending Sympiler to other matrix methods. 

\subsection{Sparse Triangular Solve}
\label{sec:trnCase}

\textbf{\textit{Theory: }} 
In the symbolic inspector, the dependency graph $DG_L$ is traversed using depth first search (DFS) to determine the inspection set for the VI-Prune transformation, which in this case is the reach-set from $DG_L$ and the right-hand side vector.
The graph $DG_L$ can also  be used to detect blocks with similar sparsity patterns, also known as supernodes, in sparse triangular solve. 
The block-set, which contains columns of $L$ grouped into supernodes, are identified by inspecting $DG_L$ using a node equivalence method. The node equivalence algorithm first assumes nodes $v_i$ and $v_j$ are equivalent and then compares their outgoing edges. If the outgoing edges go to the same destination nodes then the two nodes are equal and are merged. 


\textbf{\textit{\TransfName Transformations:}} 
Using the reach-set,VI-Prune limits the iteration spaces of the loops in triangular solve to only those that operate on the necessary nonzeros.
The VS-Block transformation changes the loops to apply blocking as shown in Figure \ref{fig:Compiler-Overview}a in triangular solve. The diagonal block of each column-block, which is a small triangular solve, is solved first. The solution of the diagonal components is then substituted in the  off-diagonal segment of the matrix.

\textbf{\textit{Symbolic Inspection:}} The time complexity of DFS on graph $DG_L$ is proportional to the number of edges traversed and the number of nonzeros in the RHS of  the system. 
The time complexity for the node equivalence algorithm is proportional to the number of nonzeros in $L$. 
We provide overheads for these methods for the tested matrices in Section \ref{sec:analysisTime}. 

\subsection{Cholesky Factorization}
\label{sec:cholCase}
\begin{figure}
\centering
\includegraphics[width=.48\textwidth]{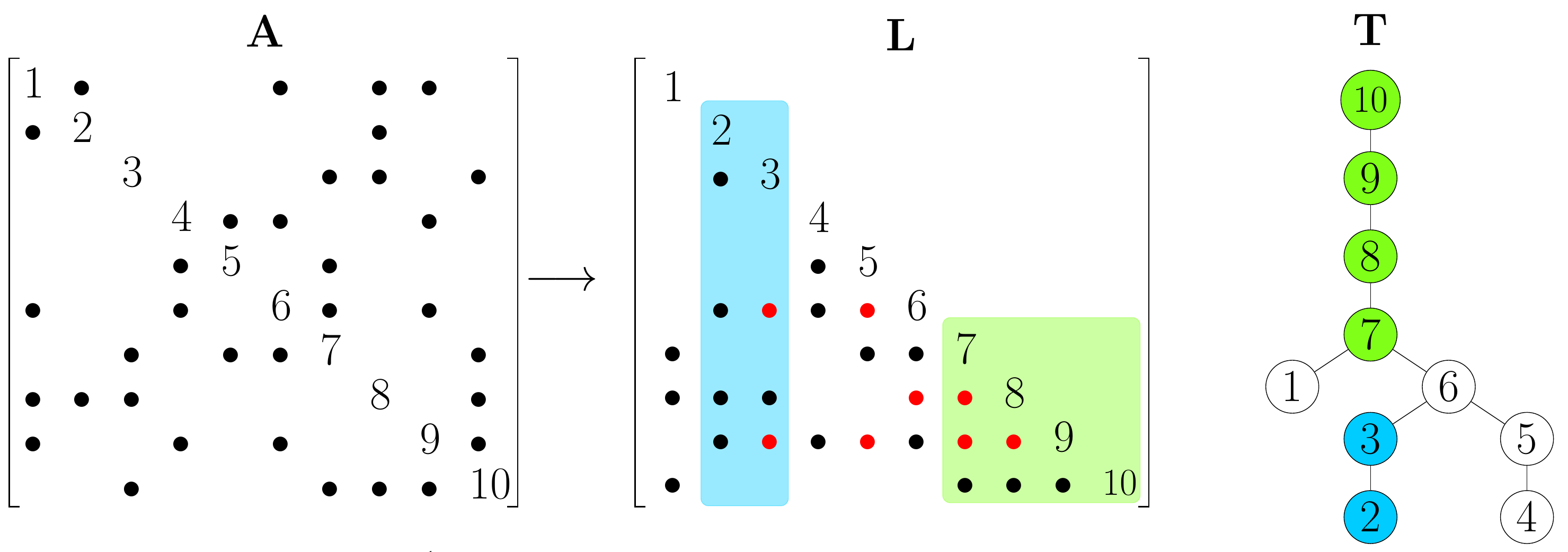}
\caption{An example matrix $A$ and its $L$ factor from Cholesky factorization. The corresponding elimination tree ($T$) of $A$ is also shown. Nodes in $T$ and columns in $L$ highlighted with the same color belong to the same supernode.  The red nonzeros in $L$ are fill-ins. }
\label{fig:etree}
 \end{figure}


Cholesky factorization is commonly used in direct solvers and is used to precondition iterative solvers. The algorithm factors a Hermitian positive definite matrix $A$ into $LL^T$, where matrix $L$ is a lower triangular matrix. Figure~\ref{fig:etree} shows an example matrix $A$ and the corresponding $L$ matrix after factorization.

\textbf{\textit{Theory: }} 
The elimination tree (etree)~\cite{davis2005row} is one of the most important graph structures used in the symbolic analysis of sparse factorization algorithms.  Figure~\ref{fig:etree} shows the corresponding elimination tree for factorizing the example matrix $A$. The etree of $A$ is a spanning tree of $G^{+}(A)$ satisfying $parent[j]= min\{i>j:L_{ij}\neq{0}\}$ where $G^{+}(A)$ is the graph of $L+L^T$. The filled graph or $G^{+}(A)$ results at the end of the elimination process and includes all edges of the original matrix $A$ as well as fill-in edges. In-depth discussions of the theory behind the elimination tree, the elimination process, and the filled graph can be found in~\cite{davis2006direct,pothen2004elimination}.

Figure~\ref{fig:codeCholesky} shows the pseudo-code of the left-looking sparse Cholesky, which is performed in two phases of \textit{update} (lines 3--6) and \textit{column factorization} (lines 7--10). The update phase gathers the contributions from the already factorized columns on the left. The column factorization phase calculates the square root of the diagonal element and applies it to the off-diagonal elements.

To find the prune-set that enables the VI-Prune transformation, the row sparsity pattern of $L$ has to be computed; Figure~\ref{fig:codeCholesky} shows how this information is used to prune the iteration space of the update phase in the Cholesky algorithm. Since $L$ is stored in column compressed format, the etree and the sparsity pattern of $A$ are used to determine the $L$ row sparsity pattern. 
A non-optimal method for finding the row sparsity pattern of row $i$ in $L$ is that for each nonzero $A_{ij}$ the etree of $A$ is traversed upwards from node $j$ until node $i$ is reached or a marked node is found. The row-count of $i$ is the visited nodes in this subtree. Sympiler uses a similar but more optimized approach from \cite{davis2006direct} to find row sparsity patterns.


Supernodes used in VS-Block for Cholesky are found with the $L$ sparsity pattern and the etree. The sparsity pattern of $L$ is different from $A$ because of fill-ins created during factorization. However, the elimination tree $T$ along with the sparsity pattern of $A$ are used to find the sparsity pattern of $L$ prior to factorization. As a result, memory for $L$ can be allocated ahead of time to eliminate the need for dynamic memory allocation. To create the supernodes, the fill-in pattern should be first determined. Equation (\ref{eq:fillin}) is based on a theorem from ~\cite{George:1981:CSL:578296} and computes the sparsity pattern of column $j$ in $L$, $L_j$, where $T(s)$ is the parent of node $s$ in $T$  and ``$\backslash$" means exclusion. The theorem states that the nonzero pattern of $L_j$ is the union of the nonzero patterns of the children of $j$ in the etree and the nonzero pattern of column $j$ in $A$.

\begin{equation}
\label{eq:fillin}
 L_j= A_j \bigcup \{j\} \bigcup \left( \bigcup\limits_{j=T(s)}  L_s \backslash \{s\} \right) \\
\end{equation} 
When the sparsity pattern of $L$ is obtained, the following rule is used to merge columns to create basic supernodes: when the number of nonzeros in two adjacent columns $j$ and $j-1$, regardless of the diagonal entry in $j-1$, is equal, and $j-1$ is the only child of $j$ in $T$, the two columns can be merged. 
%


\textbf{\textit{\TransfName transformations:}}
The VI-Prune transformation applies to the update phase of Cholesky. 
With the row sparsity pattern information, when factorizing column $i$ Sympiler only iterates over dependent columns instead of all columns smaller than $i$. The VS-Block transformation applies to both update and column factorization phases. Therefore, the outer loop in the Cholesky algorithm in Figure \ref{fig:codeCholesky} is converted to a new loop that iterates over the provided block-set. All references to the columns $j$ in the inner loops 
will be changed to the $blockSet[j]$. For the diagonal part of the column factorization, a dense Cholesky needs to be computed instead of the square root in the non-supernodal version. The resulting factor from  the diagonal elements applies to the off-diagonal rows through a sequence of dense triangular solves. VS-Block also converts the update phase from vector operations to matrix operations.   

\textbf{\textit{Symbolic Inspection:}} The computational complexity for building the etree in sympiler is nearly $O(|A|)$. 
The run-time complexity for finding the sparsity pattern of row $i$ is proportional to the number of nonzeros in row $i$ of $A$. The method is executed for all columns which results in a run-time of nearly $O(|A|)$.
The inspection overhead for finding the block-set for VS-Block includes the sparsity detection which is done in nearly $O(|A|+2n)$  and 
the supernode detection which has a run-time complexity of $O(n)$ \cite{davis2006direct}.

\subsection{Other Matrix Methods}
\label{sec:others}
The inspection graphs and inspection strategies supported in the current version of Sympiler can support a large class of commonly-used sparse matrix computations. 
The applications of the elimination tree go beyond the Cholesky factorization method and extend to some of the most commonly used sparse matrix routines in scientific applications such as LU, QR, orthogonal factorization methods~\cite{Liu:1990:RET:80034.80044},  and incomplete and factorized sparse approximate inverse preconditioner computations~\cite{janna2015use}. Inspection of the dependency graph and proposed inspection strategies that extract reach-sets and supernodes from the dependency graph are the fundamental symbolic analyses required to optimize algorithms such as rank update and rank increase methods \cite{davis2009dynamic}, incomplete LU(0) \cite{naumov2012parallel}, incomplete Cholesky preconditioners, and up-looking implementations of factorization algorithms. Thus, Sympiler with the current set of symbolic inspectors can be made to support many of these matrix methods. We plan to extend to an even larger class of matrix methods and to support more optimization methods.

%% file: SRC/Results.tex

%
%

\begin{table}[t]

\centering
 \caption{Matrix set: The matrices are sorted based on the number of nonzeros in the original matrix; nnz refers to number of nonzeros, n is the rank of the matrix.}
\begin{tabular}{|p{1cm}|c|c|p{1cm}|} 
 \hline
 Problem ID & Name & n ($10^3$) & nnz (A) ($10^6$) \\ 
 \hline
 1 & cbuckle & 13.7 & 0.677   \\
 2 & Pres\_Poisson & 14.8 & 0.716   \\
 3 & gyro & 17.4 & 1.02    \\
 4 & gyro\_k & 17.4 & 1.02    \\
 5 & Dubcova2 & 65.0 & 1.03   \\
 6 & msc23052 & 23.1 & 1.14    \\
 7 & thermomech\_dM & 204 & 1.42   \\
 8 & Dubcova3 & 147 & 3.64   \\
 9 & parabolic\_fem & 526 & 3.67  \\
 10 & ecology2 & 1000 & 5.00    \\
 11 & tmt\_sym & 727 & 5.08   \\
 

\hline
\end{tabular}
\label{tab:matrix}
\end{table}
We evaluate Sympiler by comparing the performance to two state-of-the-art libraries, namely Eigen \cite{guennebaud2010eigen} and CHOLMOD \cite{CHOLMODchen2008algorithm}, for the Cholesky factorization method and the sparse triangular solve algorithm.   
Section~\ref{Sub:meth} discusses the experimental setup and experimental methodology.
In Section~\ref{sec:symEval} we demonstrate that the transformations enabled by Sympiler generate highly-optimized codes for sparse matrix algorithms compared to state-of-the-art libraries. Although symbolic analysis is performed only once at compile-time for a fixed sparsity pattern in Sympiler, we analyze the cost of the symbolic inspector in Section~\ref{sec:analysisTime} and compare it with symbolic costs in Eigen and CHOLMOD.

\subsection{Methodology}
\label{Sub:meth}



We selected a set of symmetric positive definite matrices from~\cite{davis2011university}, which are listed in Table~\ref{tab:matrix}. The matrices originate from different domains and vary in size. All matrices have real numbers and are in double precision. The testbed architecture is a 3.30GHz Intel\textregistered Core\texttrademark i7-5820K processor with L1, L2, and L3 cache sizes of 32KB, 256KB, and 15MB respectively and turbo-boost disabled. We use OpenBLAS.0.2.19~\cite{xianyi2016openblas} for dense BLAS (Basic Linear Algebra Subprogram) routines when needed. All Sympiler-generated codes are compiled with GCC v.5.4.0 using the \texttt{-O3} option.   
Each experiment is executed 5 times and the median is reported. 


We compare the performance of the Sympiler-generated code with CHOLMOD~\cite{CHOLMODchen2008algorithm} as a specialized library for Cholesky factorization and with Eigen~\cite{guennebaud2010eigen} as a general numerical library.  CHOLMOD provides one of the fastest implementations of Cholesky factorization on single-core architectures~\cite{gould2007numerical}. Eigen supports a wide range of sparse and dense operations including sparse triangular solve and Cholesky. Thus, for Cholesky factorization we compare with both Eigen and CHOLMOD while results for triangular solve are compared to Eigen. Both libraries are installed and executed using the recommended default configuration. Since Sympiler's current version does not support node amalgamation~\cite{duff1983multifrontal}, this setting is not enabled in CHOLMOD. For the Cholesky factorization both libraries support the more commonly used left-looking (supernodal) algorithm which is also the algorithm used by Sympiler. Sympiler applies either both or one of the \transfName transformations as well as some of the enabled low-level transformations; currently, Sympiler implements unrolling, scalar replacement, and loop distibution from among the possible low-level transformations.

\subsection{Performance of Generated Code}
\label{sec:symEval}
This section shows how the combination of the introduced transformations and the decoupling strategy enable Sympiler to outperform two state-of-the-art libraries for sparse Cholesky and sparse triangular solve. 

\begin{figure}[t]
  \centering
  \includegraphics[width=\columnwidth]{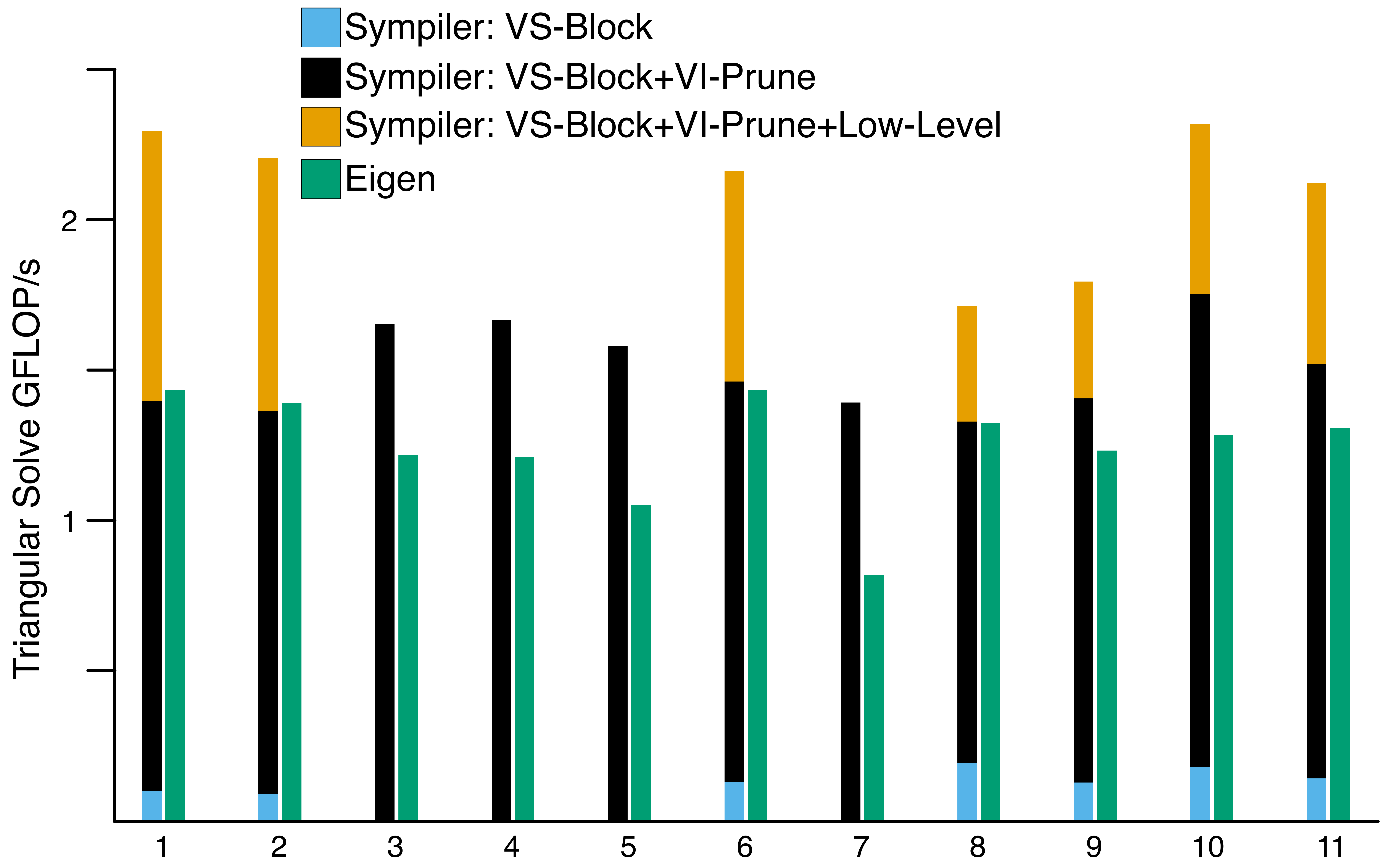}
  \caption{Sympiler's performance compared to Eigen for triangular solve. The stacked-bars show the performance of the Sympiler (numeric) code with VS-Block and VI-Prune. The effects of VS-Block, VI-Prune, and low-level transformations on Sympiler performance are shown separately.}
  \label{fig:trnsSymEig}
\end{figure}

\noindent
\textbf{Triangular solve}: Figure~\ref{fig:trnsSymEig} shows the performance of Sympiler-generated code compared to the Eigen library for a sparse triangular solve with a sparse RHS. The nonzero fill-in of the RHS in our experiments is selected to be less than 5\%. The sparse triangular system solver is often used as a sub-kernel in algorithms such as left-looking LU~\cite{davis2006direct} and Cholesky rank update methods~\cite{davis2009dynamic} or as a solver after matrix factorizations. Thus, typically the sparsity of the RHS in sparse triangular systems is close to the sparsity of the columns of a sparse matrix. For the tested problems, the number of nonzeros for all columns of $L$ was less than 5\%. 

The average improvement of Sympiler-generated code, which we refer to as Sympiler (numeric), over the Eigen library is 1.49$\times$. Eigen implements the approach demonstrated in Figure~\ref{fig:library}, where symbolic analysis is not decoupled from the numerical code. However, the Sympiler-generated code only manipulates numerical values which leads to higher performance. \autoref{fig:trnsSymEig} also shows the effect of each transformation on the overall performance of the Sympiler-generated code. In the current version of Sympiler the symbolic inspector is designed to generate sets so that VS-Block can be applied before VI-Prune. Our experiments show that this ordering often leads to better performance mainly because Sympiler supports supernodes with a full diagonal block. As support for more transformations are added to Sympiler, we will enable it to automatically decide the best transformation ordering. Whenever applicable, the vectorization and peeling low-level transformations are also applied after VS-Block and VI-Prune. Peeling leads to higher performance if applied after VS-Block where iterations related to single-column supernodes are peeled. Vectorization is always applied after VS-Block and does not lead to performance if only VI-Prune is applied.     

Matrices 3, 4, 5, and 7 do not benefit from the VS-Block transformation so their Sympiler run-times in Figure~\ref{fig:trnsSymEig} are only for VI-Prune. Since small supernodes often do not lead to better performance, Sympiler does not apply the VS-Block transformation if the average size of the participating supernodes is smaller than a threshold.
 This parameter is currently hand-tuned and is set to 160.  VS-Block is not applied to matrices 3, 4, 5, and 7 since the average supernode size is too small and thus does not improve performance. Also, since these matrices have a small column count vectorization does not payoff. 


\begin{figure}[t]
  \includegraphics[width=\columnwidth]{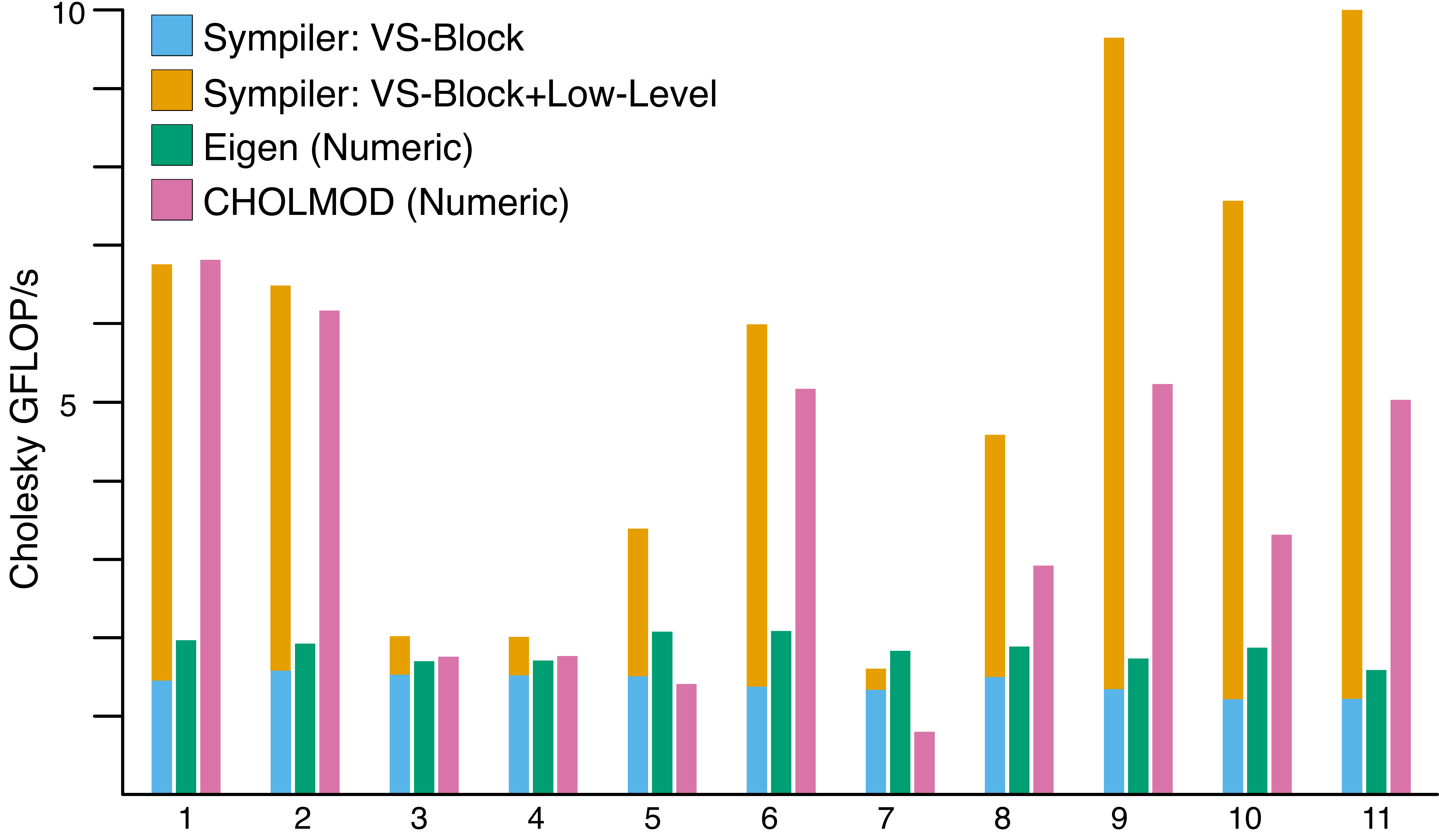}
  \caption{The performance of Sympiler (numeric) for Cholesky compared to CHOLMOD (numeric) and Eigen (numeric). The stacked-bar shows the performance of the Sympiler-generated code. The effect of VS-Block and low-level transformations are shown separately. The VI-Prune transformation is already applied to the baseline code so it is not shown here. 
  }
  \label{fig:cholFact}
\end{figure}

\noindent
\textbf{Cholesky}: 
We compare the numerical manipulation code of Eigen and CHOLMOD for Cholesky factorization with the Sympiler-generated code. The results for CHOLMOD and Eigen in Figure~\ref{fig:cholFact} refer to the numerical code performance in floating point operations per second (FLOP/s). Eigen and CHOLMOD both execute parts of the symbolic analysis only once if the user explicitly indicates that the same sparse matrix is used for subsequent executions. However, even with such an input from the user, none of the libraries fully decouple the symbolic information from the numerical code. This is because they can not afford to have a separate implementation for each sparsity pattern and also do not implement sparsity-specific optimizations. For fairness, when using Eigen and CHOLMOD we explicitly tell the library that the sparsity is fixed and thus report only the time related to the the library's numerical code (which still contains some symbolic analysis).

As shown in Figure~\ref{fig:cholFact}, for Cholesky factorization Sympiler performs up to 2.4$\times$ and 6.3$\times$ better than CHOLMOD and Eigen respectively. Eigen uses the left-looking non-supernodal approach therefore, its performance does not scale well for large matrices. CHOLMOD benefits from supernodes and thus performs well for large matrices with large supernodes. However, CHOLMOD does not perform well for some small matrices and large matrices with small supernodes. 
Sympiler provides the highest performance for almost all tested matrix types which demonstrates the power of sparsity-specific code generation.  

The application of kernel-specific and aggressive optimizations when generating code for dense sub-kernels enables Sympiler to generate fast code for any sparsity pattern. 
Since BLAS routines are not well-optimized for small dense kernels they often do not perform well for the small blocks produced when applying VS-Block to sparse codes~\cite{shin2010speeding}. Therefore, libraries such as CHOLMOD do not perform well for matrices with small supernodes. Sympiler has the luxury to generate code for its dense sub-kernels; instead of being handicapped by the performance of BLAS routines, it generates specialized and highly-efficient codes for small dense sub-kernels. If the average column-count for a matrix is below a tuned threshold, Sympiler will call BLAS routines \cite{xianyi2016openblas} instead. Since the column-count directly specifies the number of dense triangular solves, which is the most important dense sub-kernel in Cholesky, the average column-count is used to decide when to switch to BLAS routines \cite{xianyi2016openblas}. 
For example, the average column-count of matrices 3, 4, 6, and 8 are less than the column-count threshold.

%

Decoupling the prune-set calculation from the numerical manipulation phase also improves the performance of the Sympiler-generated code. As discussed in \autoref{sec:cholCase}, the sparse Cholesky implementation needs to obtain the row sparsity pattern of $L$. The elimination tree of $A$ and the upper triangular part of $A$ are both used in CHOLMOD and Eigen to find the row sparsity pattern. Since $A$ is symmetric and only its lower part is stored, both libraries compute the transpose of $A$ in the numerical code to access its upper triangular elements. Through fully decoupling symbolic analysis from the numerical code, Sympiler has the $L$ row sparsity information in the prune-set ahead of time and therefore, both the reach function and the matrix transpose operations are removed from the numeric code.



\begin{figure}[t]
  \centering
  \includegraphics[width=\columnwidth]{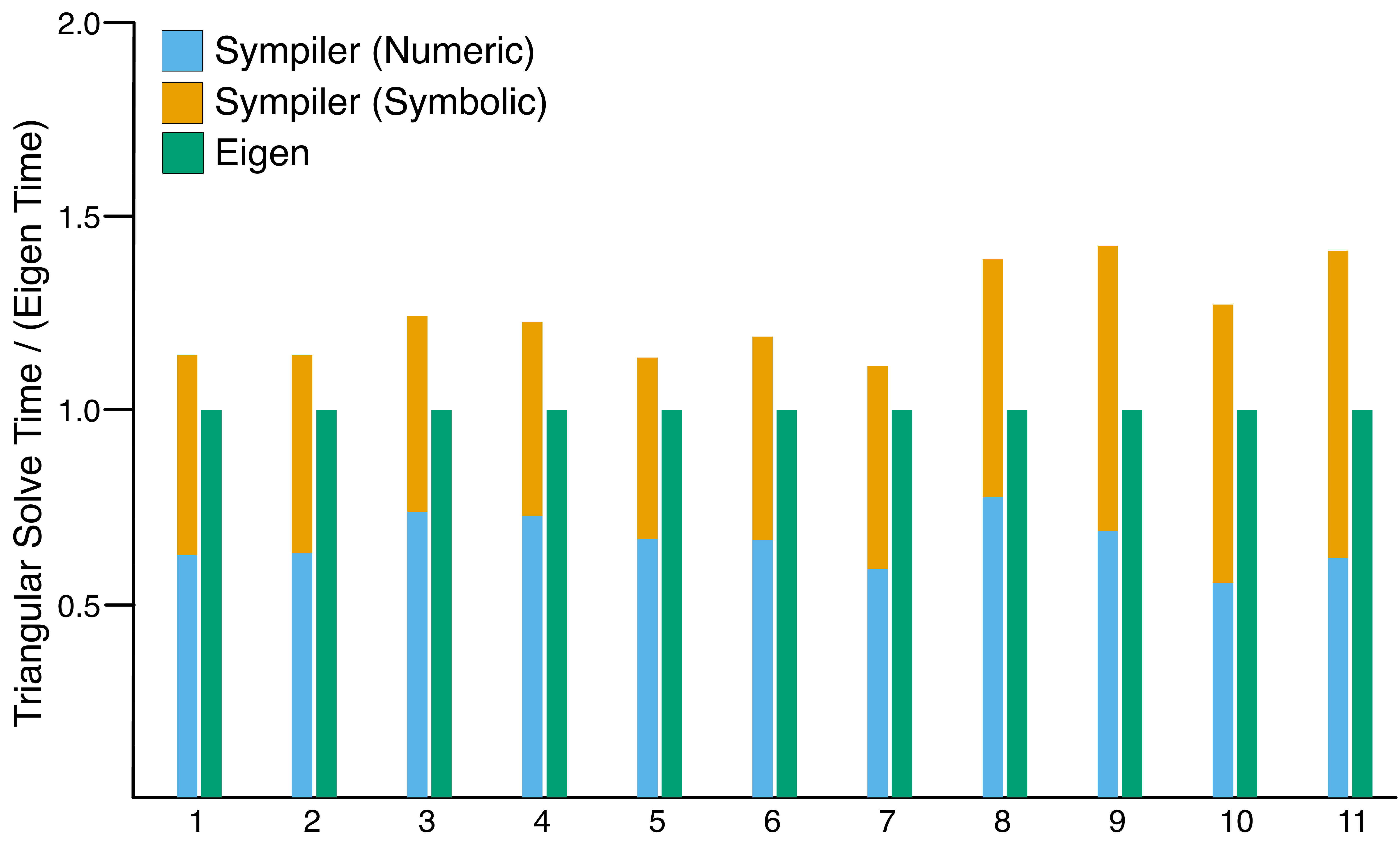}
  \caption{The figure shows the sparse triangular solve symbolic+numeric time for Sympiler and Eigen's runtime normalized over the Eigen time (lower is better).}
  \label{fig:trnsAnalysis}
\end{figure}

\subsection{Symbolic Analysis Time}
\label{sec:analysisTime}

\begin{figure}[ht]
  \centering
  \includegraphics[width=\columnwidth]{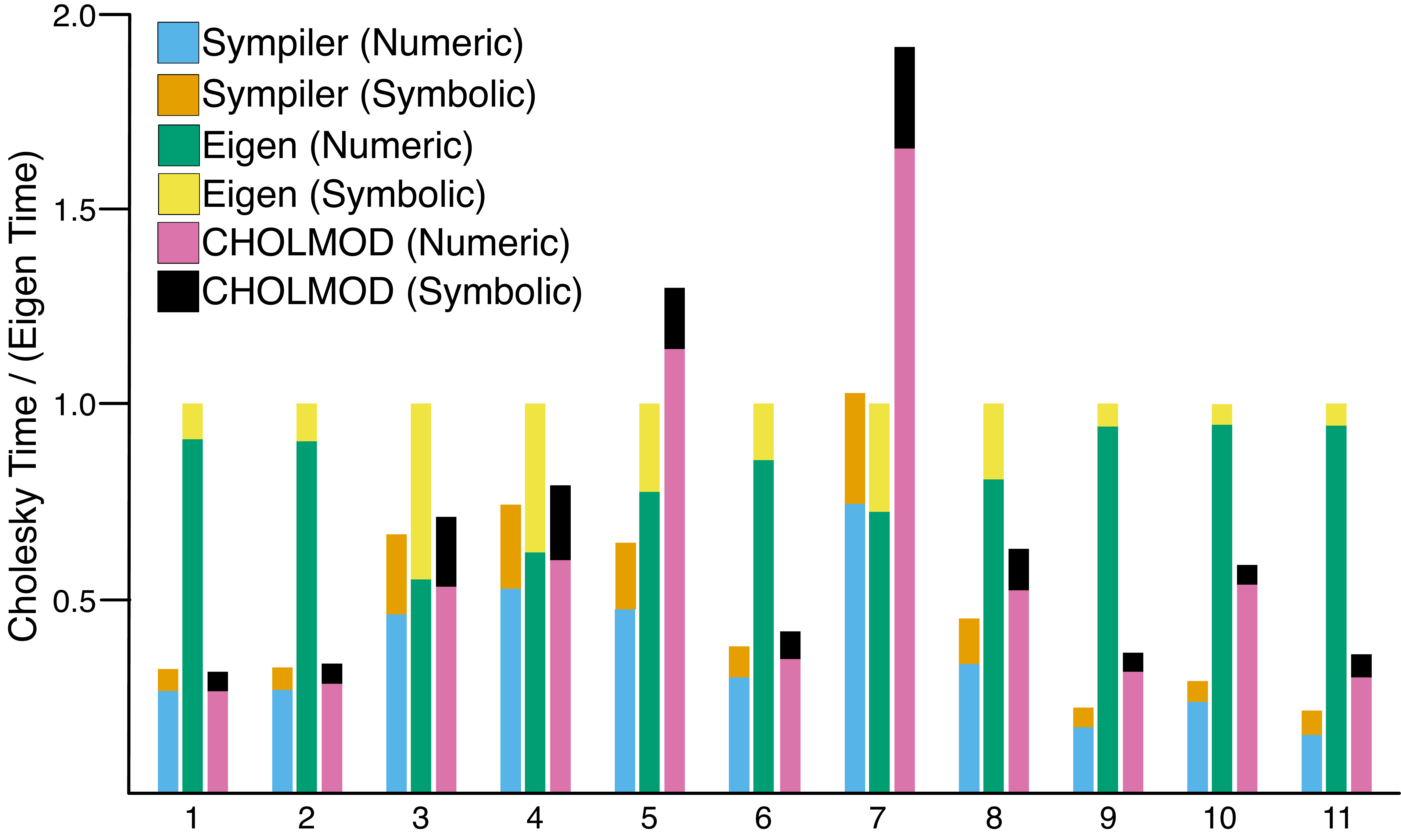}
  \caption{The figure shows the symbolic+numeric time for Sympiler, CHOLMOD, and Eigen for the Cholesky algorithm. All times are normalized over the Eigen's accumulated symbolic+numeric time (lower is better).}
  \label{fig:cholAnalys}
\end{figure}

All symbolic analysis is performed at compile-time in Sympiler and its generated code only manipulates numerical values. Since symbolic analysis is performed once for a specific sparsity pattern, its overheads amortize with repeat executions of the numerical code. However, as demonstrated in Figures~\ref{fig:trnsAnalysis} and~\ref{fig:cholAnalys} even if the numerical code is executed only once, which is not common in scientific applications, the accumulated symbolic+numeric time of Sympiler is close to Eigen for the triangular solve and faster than both Eigen and CHOLMOD for Cholesky. 

\textbf{Triangular solve}: Figure \ref{fig:trnsAnalysis} shows the time Sympiler spends to do symbolic analysis at compile-time, Sympiler (symbolic), for the sparse triangular solve. 
No symbolic time is available for Eigen since as discussed, Eigen uses the code in Figure~\ref{fig:library} for its triangular solve implementation. Figure~\ref{fig:trnsAnalysis} shows the symbolic analysis and numerical manipulation time of Sympiler normalized over Eigen's run-time. Sympiler's numeric plus symbolic time is on average 1.27$\times$ slower than the Eigen code.  In addition, code generation and compilation in Sympiler costs between 6--197$\times$ the cost of the numeric solve, depending on the matrix.  It is important to note that since the sparsity structure of the matrix in triangular solve does not change in many applications, the overhead of the symbolic inspector and compilation is only paid once. For example, in preconditioned iterative solvers a triangular system must be solved per iteration, and often the iterative solver must execute thousands of iterations ~\cite{Benzi2000robust, Papadrakakis1993accuracy, kershaw1978incomplete} until convergence since the systems in scientific applications are not necessarily well-conditioned.

\textbf{Cholesky}: 
Sparse libraries perform symbolic analysis ahead of time which can be re-used for same sparsity patterns and improves the performance of their numerical executions. We compare the analysis time of the libraries with Sympiler's symbolic inspection time.  Figure~\ref{fig:cholAnalys} provides the  symbolic analysis and numeric manipulation times for both libraries normalized to Eigen time. The time spent by Sympiler to perform symbolic analysis is referred to as Sympiler symbolic.  CHOLMOD (symbolic) and Eigen (symbolic) refer to the partially decoupled symbolic code that is only run once if the user indicates that sparsity remains static. In nearly all cases Sympiler's accumulated time is better than the other two libraries. Code generation and compilation, which are not shown in the chart, add a very small amount of time, costing at most 0.3$\times$ the cost of numeric factorization. Also, like the triangular solve example, the matrix with a fixed sparsity pattern must be factorized many times in scientific applications. For example, in Newton-Raphson (NR) solvers for nonlinear systems of equations, a Jacobian matrix is factorized in each iteration and the NR solvers require tens or hundreds of iterations to converge ~\cite{pawlowski2006globalization, de2013newton}. 

%% file: SRC/RelatedWork.tex
\label{sec:relatedWork}
\textbf{Compilers for general languages} are hampered by optimization methods that either give up on optimizing sparse codes or only apply conservative transformations that do not lead to high performance. This is due to the indirection required to index and loop over the nonzero elements of sparse data structures. 
Polyhedral methods 
  are limited when dealing with non-affine loop nests or subscripts~\cite{ancourt1991scanning,chen2012polyhedra, kelly1998optimization, quillere2000generation, vasilache2006polyhedral, tiwari2009scalable} which are common in sparse computations. 
%

To make it possible for compilers to apply more aggressive loop and data transformations to sparse codes, recent work~\cite{venkat2015loop, strout2016approach, van2010sublimation, venkat2014non, venkat2016automating} has developed compile-time techniques for automatically creating \textit{inspectors} and \textit{executors} for use at run-time. These techniques use an inspector to analyze index arrays in sparse codes at run-time and an executor that uses this run-time information to execute code with specific optimizations. These inspector-executor techniques are limited in that they only apply to sparse codes with static index arrays; such codes require the matrix structure to not change during the computation. The aforementioned approach performs well for methods such as sparse incomplete LU(0) and Gauss-Seidel methods where additional nonzeros/fill-ins are not introduced during computation. However, in a large class of sparse matrix methods, such as direct solvers including Cholesky, LU, and QR decompositions, index arrays dynamically change during computation since the algorithm itself introduces fill-ins. In addition, the indirections and dependencies in sparse direct solvers are tightly coupled with the algorithm, making it difficult to apply inspector-executor techniques. 

\textbf{Domain-specific compilers} integrate domain knowledge into the compilation process, improving the compiler's ability to transform and optimize specific kinds of computations.  Such an approach has been used successfully for stencil computations~\cite{ragan2013halide, tang2011pochoir, holewinski2012sdsl}, signal processing~\cite{pueschel2005spiral}, dense linear algebra~\cite{gunnels2001flame, spampinato2014basic}, matrix assembly and mesh analysis~\cite{alnaes2014unified,luporini2016algorithm}, simulation~\cite{kjolstad2016simit, bernstein2016ebb}, and sparse operations~\cite{davis2013algorithm,rong2016sparso}. Though the simulations and sparse compilers use some knowledge of matrix structure to optimize operations, they do not build specialized matrix solvers. 

\textbf{Specialized Libraries} are the typical approach for sparse direct solvers.  
These libraries differ in (1) which numerical methods are implemented, (2) the implementation strategy or variant of the solver, (3) the type of the platform supported, and (4) whether the algorithm is specialized for specific applications. 

Each numerical method is suitable for different classes of matrices; for example, Cholesky factorization requires the matrix be symmetric (or Hermitian) positive definite. Libraries such as SuperLU~\cite{SUdemmel1999supernodal}, KLU~\cite{KLUdavis2010algorithm}, UMFPACK~\cite{UMFPACKdavis2004algorithm}, and Eigen~\cite{guennebaud2010eigen} provide optimized implementations for LU decomposition methods. The Cholesky factorization is available through libraries such as Eigen~\cite{guennebaud2010eigen}, CSparse~\cite{davis2006direct}, CHOLMOD~\cite{CHOLMODchen2008algorithm}, MUMPS~\cite{MUMPS1amestoy2000multifrontal,MUMPS2amestoy2001fully,MUMPS3amestoy2006hybrid}, and PARDISO~\cite{PARDISIO1schenk2004solving,PARDISIO2schenk2000efficient}. QR factorization is implemented in SPARSPAK\cite{SPARSPAK1george1979design,SPARSPAK2george1981computer}, SPLOOES \cite{SPOOLESashcraft1999spooles}, Eigen~\cite{guennebaud2010eigen}, and CSparse \cite{davis2006direct}.  
 The optimizations and algorithm variants used to implement sparse matrix methods differ between libraries. For example LU decomposition can be implemented using multifrontal methods~\cite{UMFPACKdavis2004algorithm,PSPASESgupta1997highly,SuiteSparseQRdavis2011algorithm}, left-looking~\cite{SUdemmel1999supernodal,KLUdavis2010algorithm,M48duff1996design,SPARSPAK1george1979design}, right-looking~\cite{SUDISTli2003superlu_dist,S+shen2000s+,MA67duff1991factorization}, and up-looking~\cite{LDLdavis2005algorithm,NSPIVsherman1978algorithms} methods.
 Libraries are developed to support different platforms such as sequential implementations~\cite{davis2006direct,CHOLMODchen2008algorithm,KLUdavis2010algorithm}, shared memory~\cite{SuiteSparseQRdavis2011algorithm,SUMTdemmel1999asynchronous,PARDISIO1schenk2004solving}, and distributed memory~\cite{SUMTdemmel1999asynchronous,MUMPS2amestoy2001fully}.  
 Finally, some libraries are designed to perform well on matrices arising from a specific domain. For example, KLU~\cite{KLUdavis2010algorithm} works best for circuit simulation problems.  In contrast, SuperLU-MT applies optimizations with the assumption that the input matrix structure leads to large supernodes; such a strategy is a poor fit for circuit simulation problems.